\documentclass[A4,twocolumn]{article}

\usepackage{hyperref}
\usepackage{graphicx}
\author{Himel Ghosh \\ Technische Universit\"at M\"unchen}

\title{
       {\bf Enabling Efficient Serverless Inference Serving for LLM (Large Language Model) in the Cloud}
}

\begin{document}

\maketitle

\begin{abstract}
This review report discusses the cold start latency in serverless inference and existing solutions. It particularly reviews the ServerlessLLM method, a system designed to address the cold-start problem in serverless inference for large language models (LLMs). Traditional serverless approaches struggle with high latency due to the size of LLM checkpoints and the overhead of initializing GPU resources. ServerlessLLM introduces a multitier checkpoint loading system, leveraging underutilized GPU memory and storage to reduce startup times by 6–8x compared to existing methods and also proposes live inference migration and a startup-time-optimized model scheduler, ensuring efficient resource allocation and minimizing delays. This system significantly improves performance and scalability in serverless environments for LLM workloads. Besides ServerlessLLM, several other methods from recent research literature, including Rainbowcake, have been reviewed in this paper, and further discussions have been done on how the FaaS providers tackle cold starts and the possible future scopes.
\end{abstract}

\section{Introduction}
The advancement of serverless computing \cite{ref1} in cloud technology has revolutionized how computational resources are accessed, enabling developers to focus on application functionality rather than infrastructure management\cite{ref2}. In this paradigm, resources are dynamically provisioned by cloud providers, activating only when required and charging users based on utilization. However, serverless computing faces a significant hurdle when applied to Large Language Models (LLMs) \cite{ref3}, especially for applications like real-time assistants, translation tools, and chatbots that rely on swift response times. These models, due to their size—often reaching hundreds of gigabytes—and computational requirements, encounter delays due to what is known as the cold-start problem \cite{GOLEC_2024}. This latency arises when serverless functions, previously idle, initiate, leading to delays from the loading of extensive LLM checkpoints and GPU resource activation. Such cold starts can significantly hinder performance in applications requiring real-time interaction, making solutions to this problem imperative for scalable, serverless LLM deployment.

This paper investigates the current trends in tackling the cold start latency in serverless inference and specifically focusses on
ServerlessLLM by \cite{ServerlessLLM}, an innovative system tailored to reduce cold-start latency. The proposed system leverages a multi-tier checkpoint loading mechanism that optimizes GPU memory usage, alongside a live inference migration protocol and an efficient model scheduler designed to minimize startup time. ServerlessLLM enhances performance by making use of typically underutilized memory and storage resources within GPU servers, significantly reducing cold-start delays and improving cost efficiency.

This paper is structured as follows: Section 2 discusses the cold start issue in serverless LLM inference and presents an overview of current mitigation approaches. Section 3 introduces ServerlessLLM’s architecture and then reviews the methods originally introduced by \cite{ServerlessLLM}. The section also explores the necessity of live migration in LLM inference and the development of a migration strategy that ensures minimal interruption to model service. Section 4 presents the analysis of the evaluation results, demonstrating the efficacy of ServerlessLLM compared to traditional serverless frameworks. Section 5 discusses additional methods proposed by recent literature and discusses the Rainbowcake method \cite{rainbowcake} which employs a layered container architecture to reduce latency and resource wastage in serverless functions. By caching shared container layers, RainbowCake allows efficient, fine-grained resource management across functions, making it well-suited to high-concurrency environments. Section 6 offers an insight on how the Function-as-a-Service (FaaS) providers deal with cold starts. Finally, Section 7 offers potential future directions, including the integration of memory management optimizations and distributed resource allocation techniques, aiming to enhance serverless LLM scalability and efficiency further.

\begin{figure}
\centerline{
\includegraphics[width=1\columnwidth]{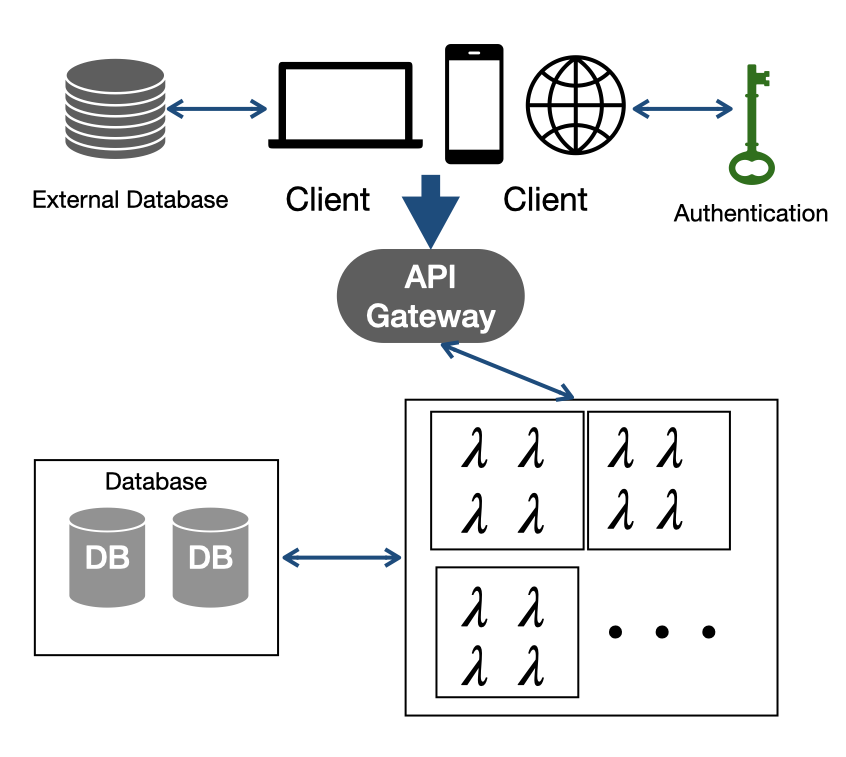}
}
\caption{Serverless Computing Architecture \cite{Ghorbian2024}}
\label{ServerlessComp}
\end{figure}

\section{Cold Start in Serverless Inference}
\label{Challenges}
Deployment of the LLMs on the serverless system often introduces significant latency due to cold-start issues. The cold start is caused by the time taken to initialize the GPUs in the serverless architecture with the LLM checkpoints, which take longer download times pertaining to their huge sizes in the hundreds of gigabytes \cite{GOLEC_2024} \cite{Ghorbian2024}. The LLM checkpoint with 130 GB (LLaMA-2-70B) from S3 or blob storage has been reported to take at least 26 seconds using a network of 5GB/s. Then the loading of these checkpoints into the GPU followed by model initialization imposes further delay. Loading the LLaMA-2-70B onto 8 GPUs takes another 84 seconds. The tokens are generated within 100ms in the LLM inference process, this loading latency significantly slows down the process.\cite{ServerlessLLM}

\begin{figure}
\centerline{
\includegraphics[width=1\columnwidth]{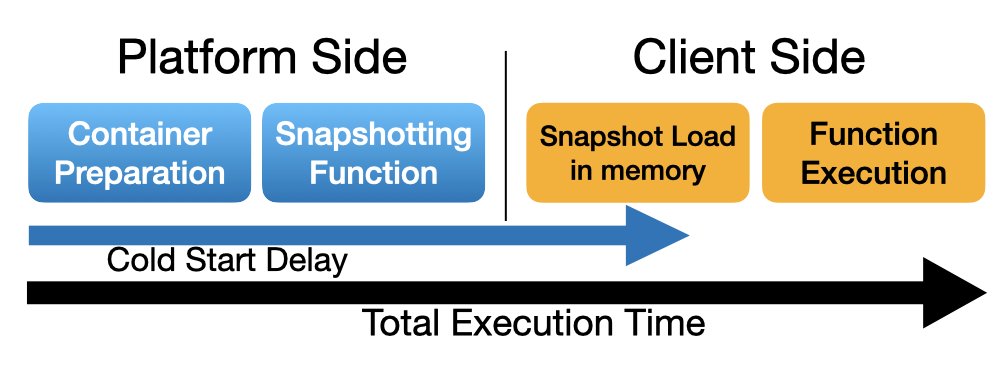}
}
\caption{Cold Start \cite{coldstart}}
\label{coldstart}
\end{figure}

\begin{figure*}[t]
\centerline{
\includegraphics[width=1\textwidth]{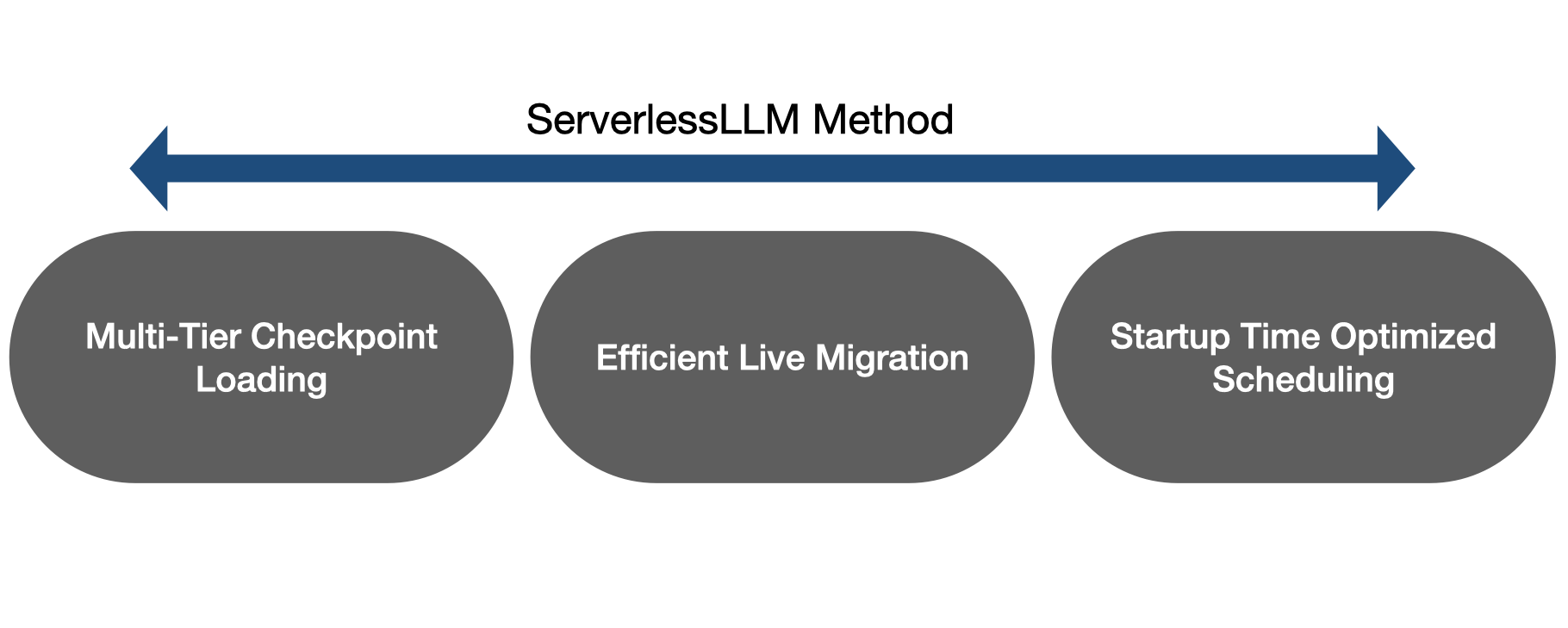}
}
\caption{Flowchart to illustrate the methods of the ServerlessLLM to mitigate Cold start problem.}
\label{Fig2}
\end{figure*}

\subsection{Cold Start mitigation approaches}
Some of the prevalent solutions to tackle the cold start issues are the following:
\begin{enumerate}
    \item Over-subscription of the GPUs to cater to peak demand scenarios. AWS serverless inference keeps a number of GPUs in the warmed state to avoid cold starts. Although it is effective for comparatively smaller models like BERT, it is not enough for LLM for their requirement of substantially high load of GPUs. \cite{Persson2023} \cite{wen2022}
    \item To avoid model downloads, model checkpoints are cached in the host memory of GPU servers. But this is effective till a few GBs and is completely inadequate for LLMs which exceeds hundreds of GB sizes. Host memory have a limited size leading to many cache misses in such cases resulting in frequent model downloads, thereby not resolving the issue. \cite{kaffes2019} \cite{du2020} \cite{anup2019} \cite{liu2023}
    \item Deployment of additional storage servers within the local cluster to cache model checkpoints, has been recommended. But this process eventually requires additional storage servers and high bandwidth network which significantly increase the costs thereby subverting the purpose of going serverless. \cite{blog1} \cite{vahi2020} \cite{zhao2024} \cite{GOLEC_2024}
    
    The use of network optimised AWS Elastic Cache servers to support a 70B model has been reported to a 100\% cost increase. \cite{ServerlessLLM}
\end{enumerate}
As is evident from the associated issues of LLM inferences in the traditional solutions, there is a strong need of new methods to tackle the cold start problem focussing the LLM requirements.

\section{Proposed Methods to mitigate Cold starts}
A new method called ServerlessLLM has been proposed by \cite{ServerlessLLM} in their paper to address this particular issue.
ServerlessLLM is a system designed specifically to reduce the cold start problem in serverless computing for Large Language Model (LLM) inference. It optimizes how models are stored, loaded, and managed, ensuring minimal startup latency.
\subsection{Design Perspective}
GPU servers are equipped with a multi-tier storage architecture offering substantial capacity and bandwidth. An 8-GPU server can handle up to 4 TB of main memory, 64 TB of NVMe SSDs, and 192 TB of SATA SSDs. However, in a serverless environment, a significant portion of the host memory and storage in these GPU servers often remains underutilized. Each GPU in these systems is linked to the host memory via a dedicated PCIe connection, ensuring a high combined bandwidth between memory and GPUs. For instance, an 8-GPU server utilizing PCIe 5.0 can deliver a 512 GBps memory-to-GPU bandwidth, with approximately 60 GBps from NVMe SSDs to memory. The suggested approach leverages the unused multi-tier storage within the server to store models locally, improving load times and reducing latency. This method proves to be cost-effective due to the reuse of underutilized resources, scalable thanks to the system’s expandable storage and bandwidth, and sustainable, as future GPU servers will offer even greater capacities and bandwidth.

\subsection{Fast Multi-tier Checkpoint Loading}
ServerlessLLM introduces a multi-tier checkpoint loading system that accelerates large model loading by leveraging a hierarchical storage structure across GPU memory, DRAM, and SSDs. This approach minimizes loading bottlenecks and improves LLM initialization times.
\begin{itemize}
    \item Loading-Optimized Checkpoint Format:
A new format facilitates sequential, chunk-based reading of model parameters, organized in partitions. This design reduces disk I/O and accelerates data transfer.
Tensors (which represent the model's weights and parameters) are organized in partitions, allowing large chunks of data to be read sequentially from storage rather than one tensor at a time. This minimizes disk I/O operations and speeds up data transfers.
    \item Multi-Tier Storage Hierarchy:
GPU Memory serve as the final destination for inference. DRAM is the Intermediate cache for quick GPU access. SSDs are Long-term checkpoint storage for faster loading compared to remote sources.
    \item Parallel Chunk-Based Loading:
Data is loaded in parallel across storage tiers, maximizing bandwidth through multi-threading and overlapping operations.
    \item Efficient Tensor Addressing: 
Each tensor's memory address is pre-computed, allowing for direct memory mapping. This eliminates unnecessary memory copy operations and further speeds up the loading process.
    \item Direct I/O and Pinned Memory:

The system uses Direct I/O (bypassing the operating system's cache) to read data directly into user space, ensuring more predictable and faster data access. Additionally, pinned memory is employed to allow direct data transfers between DRAM and GPU memory without involving the CPU, maximizing the efficiency of the PCIe bus and minimizing CPU overhead.

\end{itemize}
The system reduces cold start times significantly—up to 6X faster for small models (OPT-2.7B) and 8.2X faster for large models (LLaMA-2-70B) compared to PyTorch and SafeTensors. ServerlessLLM reaches near-maximum throughput from RAID0-NVMe SSDs, supporting predictable, high-speed data transfer ideal for real-time applications. \cite{ServerlessLLM}

\begin{figure}
\centerline{
\includegraphics[width=1\columnwidth]{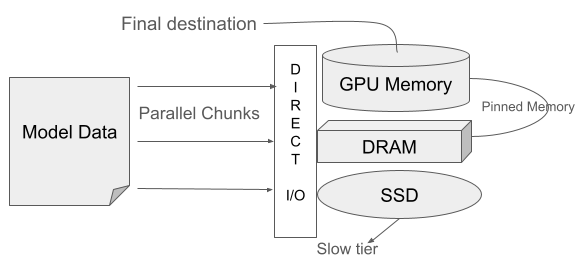}
}
\caption{The multitier storage design to mitigate cold start issues}
\label{fig3}
\end{figure}

\subsection{Review of Fast Multi-tier Checkpoint Loading}
The method has many strengths, as following:
\begin{itemize}
    \item The reduction in loading times by 8.2X speed for large models is a substantial improvement. This is particularly impressive given the size and complexity of modern LLMs, which often exceed hundreds of gigabytes in size.
    \item With the full use of the storage hierarchy and employing parallelism in data loading, the system maximizes resource utilization and reduces idle time. This makes serverless computing more resource-efficient for LLM workloads.
    \item The method is scalable, as the system is able to saturate the bandwidth of storage devices, it can handle even larger models without introducing significant additional latency.
\end{itemize}
There are also some limitations to focus on and areas for improvement:
\begin{itemize}
    \item The set of assumptions made by the authors that the checkpoints have model execution files and model parameter files, may not always hold and in that case, the architecture needs to adapt to such specific scenarios or such approaches need to be standardized.
    \item While the use of DRAM and SSDs as caching layers reduces the need for frequent accesses to slower remote storage, the system's performance is still constrained by the available memory resources. For extremely large models (e.g., those exceeding hundreds of GBs), there is a risk that the memory hierarchy could become a bottleneck, especially if multiple models are in use simultaneously.
    \item Although the method is effective in reducing latency, the use of pinned memory, Direct I/O, and parallel I/O threads may increase the overall energy consumption of the system. This is a trade-off that needs to be carefully balanced, especially for applications where power efficiency is a concern.
    \item While the system is optimized for LLM inference, it may not generalize well to other types of AI workloads that have different memory access patterns or storage requirements. Additional work may be needed to adapt the system for tasks like real-time video processing or reinforcement learning, where the data access characteristics differ from those of LLMs.
\end{itemize}
\subsection{Need for Live Migration in LLM Inference} Efficient LLM inference management is crucial when multiple models must run concurrently across servers with varied resource availability. The cold start issue, involving loading large model checkpoints into GPU memory, can significantly delay inference. We illustrate with a setup of two servers (Server 1 and Server 2) and two models (Model A and Model B), following \cite{ServerlessLLM}.

\textbf{Server conditions:}
    \begin{itemize}
        \item Server 1: Model A in DRAM, Model B in SSD, idle GPU.
        \item Server 2: Model B in DRAM, Model A running on GPU.
    \end{itemize}
\textbf{Policies:}
    \begin{itemize}
        \item GPU Availability-Driven: Model B starts on Server 1 (available GPU, but Model B is in SSD), incurring high latency to load Model B into DRAM, though Model A remains unaffected.
        
        \item Locality-Driven: Model B waits in a queue on Server 2, where it is already in DRAM, until Model A completes—delaying Model B’s startup.
       
        \item Preemption-Driven: Model A is preempted on Server 2 for Model B, then reloaded on Server 1 from DRAM/SSD, leading to significant downtime for Model A.
       
        \item Live-Migration-Supported Locality-Driven: Model A continues on Server 2 while preloading onto Server 1. Model B starts on Server 2, benefiting from DRAM locality, while Model A is seamlessly migrated to Server 1, minimizing latency for both models.
    \end{itemize}
    
This live migration approach reduces latency, optimizes resource utilization, and scales efficiently with demand and model complexity.
\begin{figure*}[t]
\centerline{
\includegraphics[width=1\textwidth]{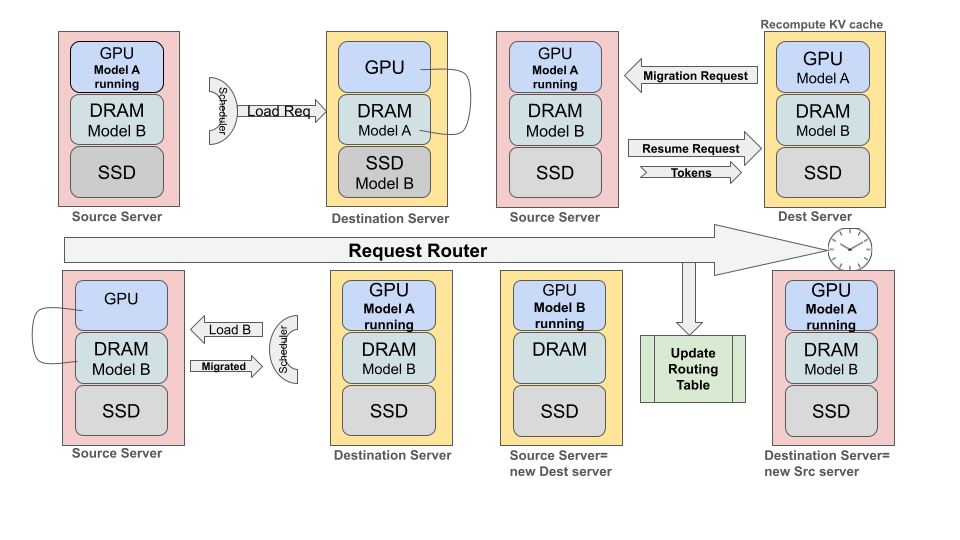}
}
\caption{Diagram that illustrates the Live Migration process.}
\label{Fig4}
\end{figure*}

\subsection{Efficient Live Migration process}
Several processes have been considered in \cite{ServerlessLLM} along with their demerits. Snapshotting LLM inference generates unnecessarily lengthy snapshots resulting in slower transfer times. Dirty-page based approach is not supported in GPU-enabled containers and VMs currently. \cite{ServerlessLLM} proposes to migrate minimal inference states to reduce network traffic. Instead of large KV-Cache (1-10s GB), migration of tokens (10-100s KB) is considered. Recomputing the KV-cache based on migrated tokens on the destination GPU is much faster. Also the destination server needs to rapidly synchronize with source server's progress causing minimal migration times. For this, KV-cache for current tokens are recomputed on the destination GPU which is significantly faster than that in the source GPU. This has been implemented as a multi-round live migration process where the destination server recomputes the KV-cache based on intermediate tokens sent by source server in each migration round and when the gap between inference states of source and destination minimises, the source stops generating and sends all tokens to the destination via request router and the further inference shifts to the destination server resulting in minimal interruption.

The process flow is as follows:
\begin{enumerate}
    \item The scheduler sends a request to the dest server to load Model A into its GPU(s). However, if an idle instance of Model A already exists on the dest server, this step is skipped, and the process moves forward without loading.
    \item Once Model A is successfully loaded, the scheduler sends a migration request to the src server. This request contains the address of the dest server, signaling that the migration process is about to begin.
    \item Upon receiving the migration request, the src server marks itself as being in a "migrating" state. If the inference on Model A is incomplete, the src server sends a resume request to the dest server, which includes the intermediate tokens (the input tokens and the output tokens generated up to this point). If the inference is already completed, the src server simply notifies the scheduler without sending the resume request.
    \item The dest server then processes the resume request by using the intermediate tokens to recompute the Key-Value (KV) cache. This step is crucial because it ensures that the dest server can seamlessly pick up the inference task from where the src server left off.
    \item Once the resume request has been processed by the dest server, the src server halts its inference operations and sends a response back to the scheduler. This response includes all tokens generated during the migration process (i.e., the intermediate tokens and any additional tokens produced during the migration), along with a flag indicating that the task has been “migrated.”
    \item The scheduler finalizes the migration by unloading Model A from the src server and initiating the loading of Model B. Simultaneously, the request router checks the flag in the inference response. If the flag indicates that the task has been migrated, the router updates its routing table to replace the src server with the dest server. It then sends all tokens to the dest server, allowing the inference task to resume seamlessly on the new server.
\end{enumerate}

\subsection{Review of Live Migration}
The autoregressive nature of LLM inference can lead to cases where a task completes on the source server (src server) before the migration fully completes. In such cases, the src server finishes the inference task and informs the request router as usual. Additionally, it notifies the loading scheduler, which then instructs the destination server (dest server) to cease resuming and terminates the migration. If the src server fails during the loading phase (before step 2), the scheduler aborts the migration and unloads the model from the dest server. If the failure occurs during migration (steps 2–3), the scheduler directs the dest server to clear any resumed KV cache and unload the model. If the dest server fails during loading, the scheduler cancels the migration.
If failure occurs while resuming, the src server notifies the scheduler of the failure and continues inference without interruption. By migrating inference tasks to servers with pre-loaded models, the process avoids the need for a complete model reload, effectively minimizing cold start times. This ensures that computational resources are more efficiently utilized, and models are able to resume inference quickly, even during migration. The ability to transfer only the token state and recompute the KV cache on the destination server allows inference to continue almost seamlessly. This ensures that the migration does not introduce significant interruptions or delays in processing, making it particularly useful for real-time LLM applications.

Although, the live migration process adds complexity to the system, particularly in terms of synchronizing between the source and destination servers. If inference completes during migration or there are timing delays, it may lead to inefficiencies or resource wastage, especially if unnecessary computations occur on the destination server. While recomputing the KV cache on the destination server reduces the need for transferring large amounts of data, it introduces computational overhead. This may still cause delays, especially for larger models, and can partially negate the latency savings from avoiding full checkpoint reloads. In case of server failures during migration, some resources may be wasted, particularly if significant amounts of computation have already occurred on the destination server before the failure is detected. Additionally, recovery from failures can introduce extra latency, affecting overall system performance.
\subsection{Startup-Time-Optimized Model Scheduling}
Another major cause of cold start latency is inefficient scheduling of inference tasks. In traditional systems, tasks are assigned to servers without considering whether the required model is already loaded. ServerlessLLM optimizes this process through startup-time-optimized model scheduling. The scheduler employs two components: model loading time estimator and model migration time estimator. 

\begin{figure}
\centerline{
\includegraphics[width=1\columnwidth]{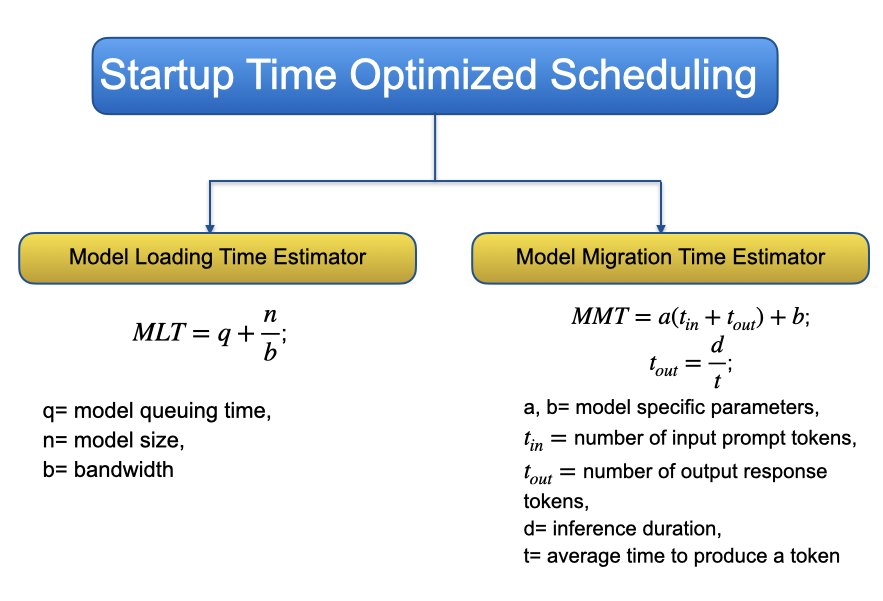}
}
\caption{Estimation times for scheduling.}
\label{fig6}
\end{figure}

\begin{itemize}
    \item Model loading Time estimation:
    The model queuing time (q), model size (n) and the bandwidth (b) are considered. ServerlessLLM tracks bandwidth for network, SSD and DRAM and then model loading time is calculated as $q + n/b$ 
    \item Model migration time estimation:
    For this, the model resuming time is calculated considering number of input prompt tokens ($t_{in}$), number of output response tokens ($t_{out}$), model specific parameters a and b that vary with each LLM batch sizes and other factors. Then the model resuming time is a x ($t_{in}$ + $t_{out}$) + b. However $t_{out}$ = d/t where d is the inference duration and t is the average time to produce a token, which are obtained by the scheduler querying the local request router.
\end{itemize}
ServerlessLLM evaluates all servers and selects the one that offers lowest estimated startup time. If no GPUs are free, the loading task is paused and retried once the request router informs the scheduler to release GPUs. \cite{ServerlessLLM} proposed a fail-safe process by utilizing a KV store to track server statuses. After server loading request, GPU status is loaded on the store, then after server confirms task confirmation, scheduler updates the storage status in the store and then notifying the request router which enables routing of the server. In failure event, latest server status from the KV store is utilized. The proposed scheduler is scalable as per current benchmarks. It is a fair process treating all models with equal priority and ensures that migrations don't impact latency.
\begin{figure*}[t]
\centerline{
\includegraphics[width=0.6\textwidth]{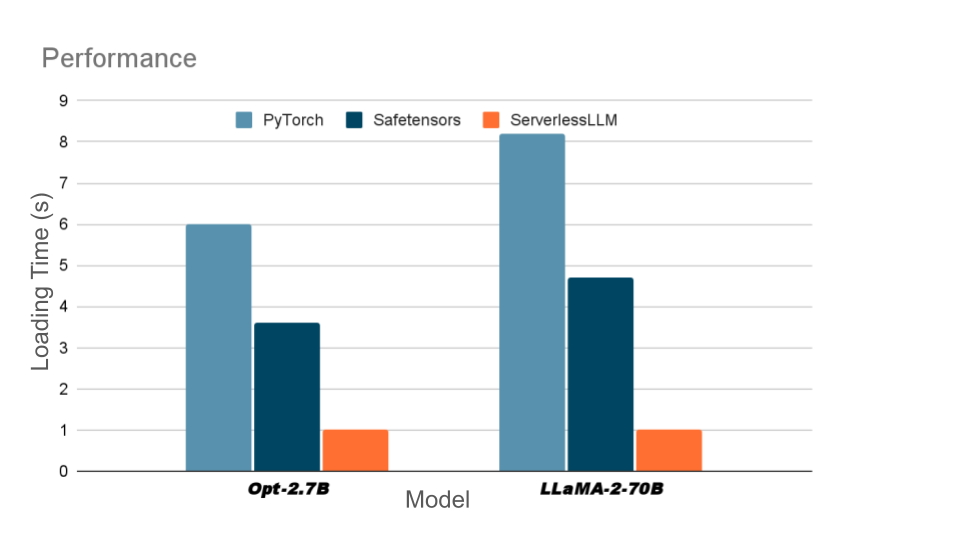}
}
\caption{Illustration of the performance of ServerlessLLM compared to baselines.}
\label{Fig5}
\end{figure*}
\section{Evaluation and Results}
 The evaluation focuses on the effectiveness of loading-optimized checkpoints, overheads from live migration, and compares ServerlessLLM against large-scale serverless workloads based on real-world data. Two main testbeds are used:
    \textbf{GPU Server}: 8 NVIDIA A5000 GPUs, 1TB DDR4 memory, and 2 AMD EPYC CPUs, connected to a MinIO storage server.
    \textbf{GPU Cluster}: 4 servers connected via 10 Gbps Ethernet, each with 4 A40 GPUs, 512 GB memory, and PCIe 4.0 NVMe SSDs.
Models like OPT, LLaMA-2, and Falcon in sizes (e.g., OPT-6.7B, OPT-13B, OPT-30B) are evaluated using datasets like GSM8K and ShareGPT for realistic input data. Workloads are derived from Azure Serverless Trace, reflecting real serverless inference conditions

\subsection{Checkpoint Loading}
ServerlessLLM achieves up to 8.2X faster model loading than PyTorch and Safetensors, notably reducing latency through chunk-based parallel loading and direct I/O. Unlike PyTorch, which copies data to host memory first, and Safetensors, which faces page faults, ServerlessLLM bypasses these overheads and consistently optimizes loading regardless of model type, focusing solely on checkpoint size. The system also loads LoRA adapters 4.4X faster and maximally utilizes storage bandwidth (e.g., RAID0-NVMe SSDs), significantly cutting cold start times in serverless environments. Bulk reading, direct I/O, multi-threading, and pinned memory collectively improve throughput by mitigating small tensor reading delays, bypassing cache, and optimizing concurrency.
\subsection{Model Scheduler}
ServerlessLLM’s cluster scheduler outperforms the default serverless scheduler and Shepherd* (a preemption-based variant), especially under high RPS conditions with the OPT-6.7B model. ServerlessLLM's locality-aware, live migration approach reduces P99 latency by 1.95X over the default scheduler, which frequently reloads models from SSD. Compared to Shepherd*, ServerlessLLM achieves 2X lower P99 latency with ShareGPT by avoiding delays from checkpoint reloads in preemption. Even at peak GPU occupancy, ServerlessLLM maintains 35–45\% lower P99 latency due to its efficient live migration, handling more migrations with minimal inference disruption.
\subsection{Entire ServerlessLLM}
In serverless deployment with Azure Serverless Trace, ServerlessLLM outperforms Ray Serve, Ray Serve with Cache, and KServe in distributed model serving. For OPT-6.7B, it initiates in 0.8 seconds, compared to 12.1 seconds (Ray Serve) and 8.2 seconds (Ray Serve with Cache). For OPT-30B, ServerlessLLM achieves a 7.5-second startup versus 213 seconds for Ray Serve, a 28X speedup. It maintains low latency across RPS ranges up to 1.1, achieving up to 212X lower latency than Ray Serve variants. Resource-efficient, ServerlessLLM reaches 4-second latency with one GPU per server, while Ray Serve with Cache needs four GPUs for 12 seconds. Even optimized, KServe trails behind in Kubernetes cluster performance.
\subsection{Remarks}
Cold-start latency poses a major challenge in serverless systems, with various approaches being employed to mitigate the issue, such as fast image pulling, lightweight isolation techniques, snapshot and restore mechanisms, resource pre-provisioning, elastic resource provisioning, and process forking. However, these methods primarily target reducing container or VM startup times without addressing the need to load large external states. Latest research has focused on optimizing cold-starts by accelerating model swaps between GPUs and host memory, but scaling these solutions for large language models (LLMs) remains a significant challenge. ServerlessLLM, in contrast, directly addresses the cold-start problem for LLM inference by introducing innovations such as optimized checkpoint formats and loading pipelines, live migration, and a cluster scheduler specifically designed for the demands of LLM workloads.

\section{Other Methods}
To address cold start delays in serverless computing, several methods have been outlined in \cite{vahi2020}, categorized broadly into approaches aimed at either minimizing the cold start duration (optimizing environments) or reducing the frequency of cold starts (pinging).

\textbf{Container Preparation Optimization:} \cite{oakes}\cite{Shillaker2018APS}\cite{lin2019}\cite{mcgrath}
\begin{enumerate}
    \item Persistent Containers: Platforms like OpenLambda and OpenWhisk pause containers after function execution to retain the container in memory, allowing rapid reuse if requests recur shortly after. However, this results in memory overhead.
    \item Container Pools: Fission and Knative platforms maintain pools of always-ready containers, which can be immediately used for new requests, significantly reducing latency but at a higher resource cost.
    \item Warmed Containers: Google Cloud Functions and AWS Lambda keep containers "warm" for a period after function execution, allowing for reuse within this timeframe if new requests arise. This approach helps reduce cold start frequency but has an associated resource wastage.
    \item Hot and Cold Queues: Some platforms utilize dual-queue architectures, with "hot" containers ready for immediate use and "cold" containers as backup. Requests are initially handled by the warm containers, switching to cold only if needed, optimizing response time at the expense of resource consumption.
    \item Application Sandboxing: The SAND platform groups functions of the same application within a single container using application-level sandboxing. By isolating functions through processes rather than containers, container preparation delay is reduced, and shared libraries load only once per container, optimizing both time and memory. \cite{ekin}
\end{enumerate}
\textbf{Library Loading Optimization:}
Frequently used libraries are preloaded into memory to avoid repetitive loading delays. This technique, similar to SAND’s approach, benefits applications where multiple functions share libraries, as they are loaded only once.

\textbf{Pinging (Reducing Cold Start Frequency)}
To reduce the frequency of cold starts, several systems implement periodic "pinging" of functions. By scheduling functions to run periodically using tools like CloudWatch, Dashbird.io, and Lambda Warmer, functions are kept "warm" by sending periodic triggers, thus preventing containers from becoming idle. \cite{cloudwatch}\cite{dashbird}\cite{lambda}

\textbf{RainbowCake} \cite{rainbowcake} is an innovative cold-start mitigation method designed for serverless computing environments, focusing on reducing latency and resource wastage during function invocations. The method employs a layered container architecture, which allows for fine-grained caching and the sharing of container layers across different functions. Specifically, it introduces three distinct layers: Bare, Lang, and User. By maintaining idle containers and enabling partial keep-alive capabilities, RainbowCake allows different functions to share compatible layers, minimizing cold-start latencies associated with serverless functions.
\cite{rainbowcake} offers significant reduction in cold-start latency through its layered caching strategy, which promotes resource efficiency by enabling container layer sharing among functions. Its lightweight design allows for scalability, effectively handling high concurrency and bursty workloads typical in serverless environments. However, the method's complexity in decision-making can hinder the optimization of trade-offs between latency and resource waste. Additionally, its reliance on accurate invocation predictions makes it sensitive to mispredictions, potentially leading to resource inefficiencies. Furthermore, while it aims to minimize operational overhead, managing multiple container layers can still introduce latency, especially in dynamic workloads.

The architecture includes two main components: the History Recorder, which captures invocation patterns using statistical modeling (e.g., Poisson distributions), and the Container Pool, responsible for managing the lifecycle of the containers based on predictive analytics of incoming requests. The dynamic scheduling of pre-warming and keep-alive strategies helps ensure that resources are allocated efficiently, addressing bursty workloads without incurring significant overhead.

\section{FaaS Providers' Approach to mitigate cold start}
\textbf{Amazon Web Services (AWS)} provides Lambda as its primary FaaS offering, which is commonly used for various serverless applications. For LLM inference, AWS Lambda is supported by additional services like SageMaker and Elastic Inference to address resource needs and cold starts. AWS introduced provisioned concurrency \cite{ref4} to pre-warm a specific number of function instances, allowing them to be instantly available when a request comes in. This significantly reduces cold start times. AWS Elastic Inference allows users to attach just the right amount of GPU to an EC2 instance or SageMaker instance. This ensures that the GPU is ready when needed, reducing cold start times for GPU-bound tasks. AWS SageMaker enables serverless inference by automatically scaling compute resources based on traffic. It also uses model caching to avoid frequent reloading of large LLM checkpoints.

\textbf{Google Cloud} offers AI Platform and Vertex AI for managing machine learning workloads. Google Cloud Functions supports the use of pre-warmed instances \cite{googlecloud} \cite{googlecloud2} \cite{googlecloud3} to mitigate cold starts. This is similar to AWS’s provisioned concurrency, where instances are kept ready to handle incoming requests without initialization delays.

For LLM inference, \textbf{Microsoft Azure} integrates with services like Azure Machine Learning and Azure Kubernetes Service (AKS) to provide GPU-enabled environments. Azure Functions' Premium Plan \cite{azure}\cite{azure2} allows for pre-warmed instances that can be kept ready to handle requests instantly, thus reducing cold start times. Azure ML provides a scalable and optimized environment for running LLMs. It offers GPU-based inference and model caching to avoid frequent cold starts. Azure Functions supports custom Docker containers that allow users to configure environments ahead of time. Pre-loading large models in custom environments can reduce cold starts by ensuring that resources are initialized and ready when the function is triggered.

One common approach by each of these FaaS providers is too keep some GPUs already in the warmed state to tackle Cold start latency. This incurs additional costs and the task becomes more expensive.

\section{Future Scopes}
The methods and evaluations presented by \cite{ServerlessLLM} demonstrate the system’s effectiveness in minimizing cold start latency. While the current approach focuses on checkpoint optimization, integrating techniques like quantization and pruning—such as 4-bit compression as seen in FlexGen \cite{clcref1}—could further reduce memory usage and loading times for larger models. These optimizations could enhance the scalability of serverless systems with minimal impact on accuracy, especially for high-throughput tasks. Additionally, adaptive resource allocation mechanisms that dynamically adjust GPU, CPU, and memory based on real-time workloads could further minimize cold start delays. Auto-scaling with resource elasticity offers another promising avenue for improving responsiveness under varying load conditions.

Exploring cross-node offloading, where model states are distributed across multiple nodes, could also mitigate cold start latencies tied to single-node limitations, particularly for large LLMs that require more memory than a single GPU node can provide. Collaborative inference—distributing tasks across multiple devices or servers—could similarly reduce the model footprint per device, addressing cold starts. Systems like Petals \cite{clcref1} illustrate the potential of such approaches, though challenges remain in balancing latency and throughput across nodes.

FastServe \cite{clcref2} demonstrates the potential of preemptive scheduling at the token level to reduce job completion times (JCT) during LLM inference. Extending these strategies to optimize first-token latency while keeping overhead low could further enhance cold-start efficiency in serverless LLM deployments. Current key-value cache management methods introduce overhead due to memory constraints, and innovations in memory-swapping techniques or distributed memory architectures could optimize resource utilization, reducing cold starts when memory is limited.

Systems like Tabi \cite{clcref3}, which employ a multilevel inference engine that escalates from smaller models to LLMs only when necessary, offer a promising solution. Adapting this approach to serverless environments could preemptively handle simpler queries with lightweight models, reserving larger LLMs for more complex tasks, thus minimizing cold start latencies. Future research could focus on developing adaptive scheduling systems that automatically select the appropriate model based on query complexity, avoiding the use of heavy LLMs for tasks that require far less computation. Attention-based pruning techniques, as demonstrated by Tabi \cite{clcref3}, could also be explored further to reduce computational overhead in serverless settings, enhancing responsiveness by minimizing model loading times and cold start delays.

ZeRO-Infinity’s \cite{clcref4} memory-centric tiling and offload engine provide a compelling solution for efficiently utilizing heterogeneous memory systems, such as GPU, CPU, and NVMe. Applying similar strategies to serverless LLM inference could reduce cold start times by offloading less frequently accessed data to NVMe or CPU, easing GPU memory pressure. ZeRO-Infinity’s dynamic prefetching and overlap-centric design \cite{clcref4}, which overlaps data movement with computation, could also be implemented in serverless platforms to reduce wait times and ensure more seamless inference during high-demand scenarios. As serverless LLM workloads scale to trillions of parameters, techniques from ZeRO-Infinity—like NVMe offload and superlinear scalability—can help maintain performance while mitigating cold start issues.

\bibliographystyle{plain}
\bibliography{seminarpaper}
\end{document}